# Band structure engineering in $(Bi_{1-x}Sb_x)_2Te_3$ ternary topological insulators

Jinsong Zhang[1,*], Cui-Zu Chang[1,2*], Zuocheng Zhang[1], Jing Wen[1], Xiao Feng[2], Kang Li[2], Minhao Liu[1], Ke He[2,†], Lili Wang[2], Xi Chen[1], Qi-Kun Xue[1,2], Xucun Ma[2] & Yayu Wang[1,†]

[1]*State Key Laboratory of Low Dimensional Quantum Physics, Department of Physics, Tsinghua University, Beijing 100084, P. R. China*

[2]*Institute of Physics, Chinese Academy of Sciences, Beijing 100190, P. R. China*

*\* These authors contributed equally to this work.*

[†] Emails: kehe@aphy.iphy.ac.cn; yayuwang@tsinghua.edu.cn

**Topological insulators (TI) are novel quantum materials with insulating bulk and topologically protected metallic surfaces with Dirac-like band structure. The most challenging problem facing the current TI materials is the existence of significant bulk conduction. Here we show the band structure engineering of TIs by molecular beam epitaxy growth of $(Bi_{1-x}Sb_x)_2Te_3$ ternary compounds. The topological surface states are shown to exist over the entire composition range of $(Bi_{1-x}Sb_x)_2Te_3$, indicating the robustness of bulk $Z_2$ topology. Most remarkably, the band engineering leads to ideal TIs with truly insulating bulk and tunable surface states across the Dirac point that behave like one quarter of graphene. This work demonstrates a new route to achieving intrinsic quantum transport of the topological surface states and designing conceptually new TI devices with well-established semiconductor technology.**

**Introduction**

The topological surface states of three dimensional TIs possess a single spin-polarized Dirac cone originated from strong spin-orbit coupling[1-3]. The unique surface states are expected to host exotic topological quantum effects[4-6] and find applications in spintronics and quantum computation. The experimental realization of these ideas requires fabrication of versatile devices based on bulk-insulating TIs with tunable surface states. However, the currently available TI materials exemplified by $Bi_2Se_3$ and $Bi_2Te_3$[7] always show conductive bulk states due to the defect-induced charge carriers. Tuning the band structure of the TIs to eliminate the bulk states is one of the most urgent tasks in the field, but the problem remains unsolved despite extensive efforts involving nanostructuring[8], chemical doping[9-15] and electrical gating[16-19].

Energy band engineering in conventional semiconductor is a powerful approach for tailoring the electronic structure of materials[20]. A notable example is the isostructural isovalent alloy of the III-V semiconductors $Al_xGa_{1-x}As$ grown on GaAs by MBE, in which the energy gap can be tuned continuously by the mixing ratio of AlAs and GaAs. Such energy band tuning has been essential for heterostructures which were later used for discovery of fractional quantum Hall effect and invention of high speed electronics.

Inspired by this idea, we conceive a new route for engineering the band structure of TIs by fabricating alloys of $Bi_2Te_3$ and $Sb_2Te_3$. Both TIs are V-VI compounds with the same crystal structure and close lattice constants[7], making it ideal to form $(Bi_{1-x}Sb_x)_2Te_3$ ternary compounds with arbitrary mixing ratio and negligible strain (Fig. 1a). The potential advantages of mixing the two TIs can be anticipated from their complementary electronic properties. Figure 1b illustrates the band structure of pure $Bi_2Te_3$[7,10,21], which reveals two major drawbacks of the surface Dirac band in $Bi_2Te_3$. First, the Dirac point (DP) is buried in the bulk valence band (BVB) hence cannot be accessed by transport experiment and, more seriously, the Fermi level ($E_F$) lies in the bulk conduction band (BCB) due to the electron-type bulk carriers induced by Te vacancies. The band structure of pure $Sb_2Te_3$[7,21], on the other hand, is drastically different. As shown schematically in Figure 1c, here the DP lies within the bulk gap and the $E_F$ lies in the BVB due to the hole-type bulk carriers induced by Sb-Te antisite defects. Intuitively, by mixing the two compounds one can simultaneously achieve charge compensation and tune the position of the DP, which may lead eventually to an ideal TI with exposed DP and insulating bulk.

Here we report the band structure engineering in TIs by fabricating alloys of $Bi_2Te_3$ and $Sb_2Te_3$ using state-of-the-art molecular beam epitaxy (MBE). Transport and angle resolved photoemission spectroscopy (ARPES) measurements show that the band engineering technique allows us to achieve ideal TIs with truly insulating bulk. The surface states can be tuned systematically across the Dirac point and the transport properties are consistent with that of a single spin-polarized Dirac cone.

## Results

**Sample structure.** During the MBE growth of the $(Bi_{1-x}Sb_x)_2Te_3$ films, the growth rate is calibrated by a real-time RHEED (Reflection High Energy Electron Diffraction) intensity oscillation measured on the (00) diffraction. Supplementary Figure S1 shows a typical 1×1 RHEED pattern taken on a $(Bi_{1-x}Sb_x)_2Te_3$ film with five quintuple layers (QL) thickness. The sharpness of the feature provides a clear evidence for the high quality of the sample. The 5QL thickness is used for all $(Bi_{1-x}Sb_x)_2Te_3$ films studied in this work because in this ultrathin regime the surface states dominate charge transport, and meanwhile the films are thick enough that the top and bottom surfaces are completely decoupled. Further discussion about the film thickness issue can be found in the Supplementary Information.

**Electronic structure.** The electronic structures of the $(Bi_{1-x}Sb_x)_2Te_3$ films are measured by ARPES on a sample setup as illustrated in Supplementary Figure S2. The ARPES band maps of eight $(Bi_{1-x}Sb_x)_2Te_3$ films with $0 \leq x \leq 1$ are shown in Figures 2a to 2h. The pure $Bi_2Te_3$ film shows well-defined surface states with massless Dirac-like dispersion (Fig. 2a), similar to that of the cleaved $Bi_2Te_3$ crystal[10]. With the addition of Sb, the Dirac-like topological surface states can be clearly observed in all $(Bi_{1-x}Sb_x)_2Te_3$ films from $x = 0$ to 1, while the Dirac cone geometry changes systematically. With increasing $x$, the slope of the Dirac line shape becomes steeper, indicating an increase of the Dirac fermion velocity $v_D$ defined by the linear dispersion $\varepsilon = v_D \cdot \hbar k$ near the DP. Meanwhile, the $E_F$ moves downwards from the BCB, indicating the reduction of the electron-type bulk carriers. Moreover, the DP moves upwards relative to the BVB due to the increasing weight of the $Sb_2Te_3$ band structure. When the Sb content is increased to $x = 0.88$ (Fig. 2e), both the DP and $E_F$ lie within the bulk energy gap. The system is now an ideal TI with a truly insulating bulk and a nearly symmetric surface Dirac cone with exposed DP. Notably, when $x$ increases from $x = 0.94$ (Fig. 2f) to 0.96 (Fig. 2g), $E_F$ moves from above the DP to below it, indicating a crossover from electron- to hole-type Dirac fermion gas. The charge neutrality point (CNP) where $E_F$ meets DP can thus be identified to be located between $x = 0.94$ and 0.96.

It is quite remarkable that the topological surface states exist in the entire composition range of $(Bi_{1-x}Sb_x)_2Te_3$, which implies that the nontrivial $Z_2$ topology of the bulk band is very robust against alloying. This is in contrast to the $Bi_{1-x}Sb_x$ alloy, the first discovered 3D TI in which the topological surface states only exist within a narrow composition range near $x = 0.10$[22,23].

Figures 3a to 3c summarize the characteristics of the surface Dirac band in the $(Bi_{1-x}Sb_x)_2Te_3$ compounds, which are extracted following the procedure presented in the Supplementary Information and illustrated in Supplementary Figures S3 and S4. The position of the DP rises continuously from below the top of BVB near the $\Gamma$ point at $x = 0$ to way above that at $x = 1$ (Fig. 3a). This is accompanied by a drastic change of the relative position of $E_F$ and DP (Fig. 3b), which determines the type and density of Dirac fermions. Furthermore, $v_D$ increases from $3.3\times10^5$m/s at $x = 0$ to $4.1\times10^5$m/s at $x = 1$ (Fig. 3c). Since the three defining properties of the Dirac cone are systematically varied between that of pure $Bi_2Te_3$ and $Sb_2Te_3$, the $(Bi_{1-x}Sb_x)_2Te_3$ ternary compounds are effectively a series of new TIs. The bulk electronic structures, including the geometry of BCB and BVB as well as the energy gap between them, are also expected to change with $x$. They are of interests in their own rights, but will not be the main focus of the current work.

**Transport properties.** The systematic Dirac band evolution also manifests itself in the transport properties. Figure 4 displays the variation of two dimensional (2D) sheet resistance ($R_\square$) with temperature ($T$) for eight 5QL $(Bi_{1-x}Sb_x)_2Te_3$ films with $0 \leq x \leq 1$. In pure $Bi_2Te_3$ the resistance shows metallic behavior at high $T$ and becomes weakly insulating at very low $T$. With increasing $x$, the $R_\square$ value keeps rising and the insulating tendency becomes stronger, reflecting the depletion of electron-type bulk carriers and surface Dirac fermions. At $x = 0.94$ when $E_F$ lies just above DP, the resistance reaches the maximum value and shows insulating behavior over the whole $T$ range. With further increase of Sb content from $x = 0.96$ to 1, the resistance decreases systematically because now $E_F$ passes DP and more hole-type carriers start to populate the surface Dirac band. The high $T$ metallic behavior is recovered in pure $Sb_2Te_3$ when the hole-type carrier density becomes sufficiently high.

Figure 5a displays the variation of the Hall resistance ($R_{yx}$) with magnetic field ($H$) measured on the 5QL $(Bi_{1-x}Sb_x)_2Te_3$ films at $T = 1.5$K. For films with $x \leq 0.94$, the $R_{yx}$ value is always negative, indicating the existence of electron-type carriers. The weak-field slope of the Hall curves, or the Hall coefficient $R_H$, increases systematically with $x$ in this regime. Since the 2D carrier density $n_{2D}$ can be derived from $R_H$ as $n_{2D} = 1/eR_H$ ($e$ is the elementary charge), this trend confirms the decrease of electron-type carrier density with Sb doping. As $x$ increases slightly from 0.94 to 0.96, the Hall curve suddenly jumps to the positive side with a very large slope, which indicates the reversal to hole-type Dirac fermions with a small carrier density. At even higher $x$, the slope of the positive curves decreases systematically due to the increase of hole-type carrier density.

The evolution of the Hall effect is totally consistent with the surface band structure revealed by ARPES in Figure 2. To make a more quantitative comparison between the two experiments, we use the $n_{2D}$ derived from the Hall effect to estimate the Fermi wavevector $k_F$ of the surface Dirac band. By assuming zero bulk contribution and an isotropic circular Dirac cone structure (Fig. 5b), $k_F$ can be expressed as

$$\frac{D \cdot k_F^2}{4\pi} = |n_{SS}|. \qquad [1]$$

Here $D$ is the degeneracy of the Dirac fermion and $|n_{SS}| = \frac{1}{2}|n_{2D}|$ is the carrier density per surface if we assume that the top and bottom surfaces are equivalent. Figure 5c shows that when we choose $D = 1$, the $k_F$ values derived from the Hall effect match very well with that directly measured by ARPES. This remarkable agreement suggests that the transport properties of the TI surfaces are consistent with that of a single spin-polarized Dirac cone, or a quarter of graphene, as expected by theory.

Figures 5d to 5f summarize the evolution of the low $T$ transport properties with Sb content $x$. The resistance value shows a maximum at $x = 0.94$ with $R_\square > 10$ k$\Omega$ and decreases systematically on both sides. Correspondingly, the carrier density $|n_{2D}|$ reaches a minimum at $x = 0.96$ with $|n_{2D}| = 1.4\times10^{12}/\text{cm}^2$ and increases on both sides. Using the measured $R_\square$ and $|n_{2D}|$, the mobility $\mu$ of the Dirac fermions can be estimated by using the Drude formula $\sigma_{2D} = |n_{2D}|e\mu$, where $\sigma_{2D} = 1/R_\square$. As a function of $x$ the mobility also peaks near the CNP and decreases rapidly on both sides. The "V"-shaped dependence of the transport properties on the Sb content $x$ clearly demonstrate the systematic tuning of the surface band structure across the CNP.

## Discussion

The good agreement with ARPES suggests that the transport results are consistent with the properties of the surface Dirac fermions without bulk contribution. Moreover, the alloying allows us to approach the close vicinity of the CNP, which gives a very low $|n_{2D}|$ in the order of $1\times10^{12}/\text{cm}^2$. The $(\text{Bi}_{1-x}\text{Sb}_x)_2\text{Te}_3$ compounds thus represent an ideal TI system to reach the extreme quantum regime because now a strong magnetic field can squeeze the Dirac fermions to the lowest few Landau levels. Indeed, the Hall resistance of the $x = 0.96$ film shown in Figure 5a is close to 7k$\Omega$ at 15T, which is a significant fraction of the quantum resistance. Future transport measurements on $(\text{Bi}_{1-x}\text{Sb}_x)_2\text{Te}_3$ films with higher mobility to even stronger magnetic field hold great promises for uncovering the unconventional quantum Hall effect of the topological surface states[24,25].

The band structure engineering offers many enticing opportunities for designing conceptually new experimental or device schemes based on the TIs. For example, we can apply the idea of compositionally graded doping (CGD) in conventional semiconductor devices[20] to the TIs to achieve spatially variable Dirac cone structures. Figure 6a illustrates the schematic of vertical CGD TIs, in which the top and bottom surfaces have opposite types of Dirac fermions and can be used for studying the proposed topological exciton condensation[26]. The spatial asymmetry of the surface Dirac bands can also be used to realize the electrical control of spin current by using the spin-momentum locking in the topological surfaces for spintronic applications[27]. Figure 6b illustrates the schematic of horizontal CGD TIs, by which a topological *p-n* junction between hole- and electron-type TIs can be fabricated.

## Methods

**MBE sample growth.** The MBE growth of TI films on insulating substrate has been reported before by the same group[28]. The $(Bi_{1-x}Sb_x)_2Te_3$ films studied here are grown on sapphire (0001) in an ultrahigh vacuum (UHV) MBE-ARPES-STM combined system with a base pressure of $1\times10^{-10}$ Torr. Prior to sample growth, the sapphire substrates are first degassed at 650°C for 90 minutes and then heated at 850°C for 30 minutes. High purity Bi (99.9999%), Sb (99.9999%), and Te (99.999%) are evaporated from standard Knudsen cells. In order to reduce Te vacancies, the growth is kept in Te-rich condition with the substrate temperature at 180°C. The Bi:Sb ratio is controlled by the temperatures of the Bi and Sb Knudsen cells. The $x$ value in the $(Bi_{1-x}Sb_x)_2Te_3$ film is determined through two independent methods, as discussed in detail in the Supplementary Information.

**ARPES measurements.** The *in situ* ARPES measurements are carried out at room temperature by using a Scienta SES2002 electron energy analyzer. A Helium discharge lamp with a photon energy of $hv = 21.218$ eV is used as the photon source. The energy resolution of the electron energy analyzer is set at 15meV. All the spectra shown in the paper are taken along the K-Γ-K direction. To avoid sample charging during ARPES measurements due to the insulating sapphire substrate, a 300-nm-thick titanium (Ti) film is deposited at both ends of the substrate which is connected to the sample holder. The sample is grounded through these contacts once a continuous film is formed. The sample setup for the ARPES measurements is illustrated schematically in the Supplementary Figure S2.

**Transport measurements.** The transport measurements are performed *ex situ* on the 5QL $(Bi_{1-x}Sb_x)_2Te_3$ films grown on sapphire (0001) substrate. To avoid possible contamination of the TI films, a 20nm-thick amorphous Te capping layer is deposited on top of the films before we take them out of the UHV growth chamber for transport measurements. The Hall effect and resistance are measured using standard ac lock-in method with the current flowing in the film plane and the magnetic field applied perpendicular to the plane. The schematic device setup for the transport measurements is shown in Supplementary Figure S5. The 20nm amorphous Te capping layer causes no significant change of the TI surface electronic structure and makes negligible contribution to the total transport signal, as shown in Supplementary Figure S6 and discussed in the Supplementary Information.

**Acknowledgments** We acknowledge S. C. Zhang and Y. B. Zhang for suggestions and comments. This work was supported by the National Natural Science Foundation of China, the Ministry of Science and Technology of China (grant number 2009CB929400), and the Chinese Academy of Sciences.

**Author contributions** K.H., Y.W., X.C.M. and Q.K.X. designed the research. C.Z.C., J.W., X.F. and K.L. carried out the MBE growth of the samples and ARPES measurements. J.S.Z., Z.C.Z and M.H.L. carried out the transport measurements. L.L.W., X.C., and X.C.M. assisted in the experiments. K.H., Y.W. and Q.K.X. prepared the manuscript. All authors have read and approved the final version of the manuscript.

**Competing financial interests** The authors declare no competing financial interests.

**Author Information** Correspondence and requests for materials should be addressed to K.H. (kehe@aphy.iphy.ac.cn) or Y.W. (yayuwang@tsinghua.edu.cn).

**Figure 1 | The schematic crystal and electronic structures of the $(Bi_{1-x}Sb_x)_2Te_3$ compounds. a**, The tetradymite-type crystal structure of $(Bi_{1-x}Sb_x)_2Te_3$ where the Bi atoms are partially substituted by Sb. **b**, The schematic electronic band structure of pure $Bi_2Te_3$ and (**c**) pure $Sb_2Te_3$ based on theoretical calculations[7] and ARPES experiments[10,21].

**Figure 2 | ARPES results on the 5QL $(Bi_{1-x}Sb_x)_2Te_3$ films measured along the K-Γ-K direction.** From **a** to **h**, the measured band structures of $(Bi_{1-x}Sb_x)_2Te_3$ films with $x = 0$, 0.25, 0.62, 0.75, 0.88, 0.94, 0.96 and 1.0 respectively. The Dirac-like topological surface states exist in all films. The yellow dashed line indicates the position of the Fermi level ($E_F$). The blue and red dashed lines indicate the Dirac surface states with opposite spin polarities and they intersect at the Dirac point (DP).

**Figure 3 | Evolution of the surface band characteristics with $x$ obtained from the ARPES data in $(Bi_{1-x}Sb_x)_2Te_3$. a**, Relative position (or energy difference) between the DP and the top of BVB near the Γ point. **b**, Relative position between the DP and the $E_F$. **c**, The Dirac fermion velocity $v_D$ ($v_D \sim \tan\theta$) extracted from the linear dispersion near the DP. All three quantities evolve smoothly from that of pure $Bi_2Te_3$ ($x = 0$) to pure $Sb_2Te_3$ ($x = 1$).

**Figure 4 | Two dimensional (2D) sheet resistance ($R_\square$) versus temperature ($T$) for eight 5QL $(Bi_{1-x}Sb_x)_2Te_3$ films.** $R_\square$ value keeps rising and the insulating tendency becomes stronger with increasing Sb content from $x = 0$ to 0.94 due to the reduction of electron-type carriers. From $x = 0.96$ to 1 the trend is reversed, i.e., $R_\square$ value decreases

and the insulating tendency becomes weaker with increasing Sb content due to the increasing density of hole-type carriers.

**Figure 5 | The Hall effect and summary of the transport results. a**, The field dependence of the Hall resistance $R_{yx}$ for the eight $(Bi_{1-x}Sb_x)_2Te_3$ films measured at $T$ = 1.5K. From top to bottom, the curves are the Hall traces of $(Bi_{1-x}Sb_x)_2Te_3$ films with $x$ = 0.96, 0.98, 1.0, 0, 0.50, 0.75, 0.88 and 0.94 respectively. The evolution of the Hall effect reveals the depletion of electron-type carriers (from $x$ = 0 to 0.94), the reversal of carrier type (from $x$ = 0.94 to 0.96), and the increase of hole-type carrier density (from $x$ = 0.96 to 1.0). **b**, Schematic sketch of an isotropic circular Dirac cone where the Fermi wavevectors $k_F$ is marked. The blue arrows indicate the helical spin texture. **c**, The $k_F$ of the Dirac cone derived from the Hall effect (black open squares) agree well with that directly measured by ARPES (red solid circles) if we assume a single spin-polarized Dirac cone on each surface. The $k_F$ is defined to be negative for hole-type Dirac fermions. The sheet resistance $R_\square$ (**d**), the carrier density $|n_{2D}|$ (**e**) and the mobility $\mu$ of the Dirac fermions (**f**) measured at $T$ = 1.5K all show "V"-shaped $x$ dependence near the charge neutrality point (CNP).

**Figure 6 | Schematic device structures of spatially variable Dirac bands grown by compositionally graded doping (CGD) of $(Bi_{1-x}Sb_x)_2Te_3$ films.** Vertical CGD TIs (**a**) is an ideal system for studying the topological exciton condensation and electrical control of spin current. Horizontal CGD TIs (**b**) can be used to fabricate a topological *p-n* junction.

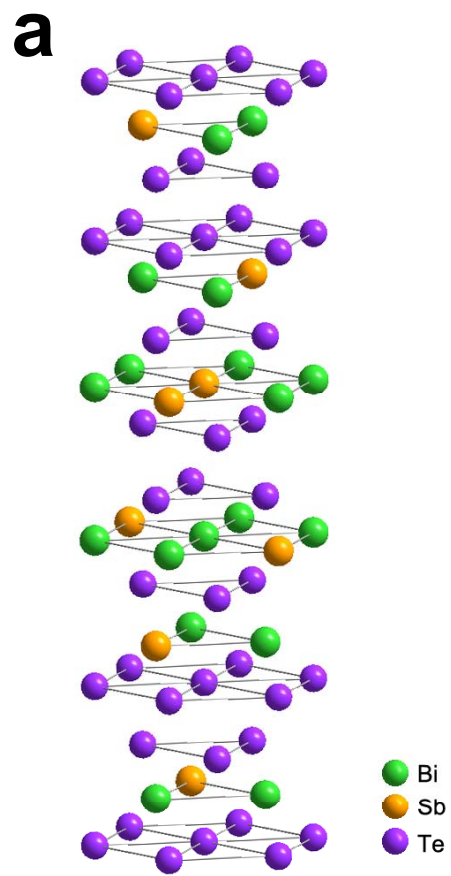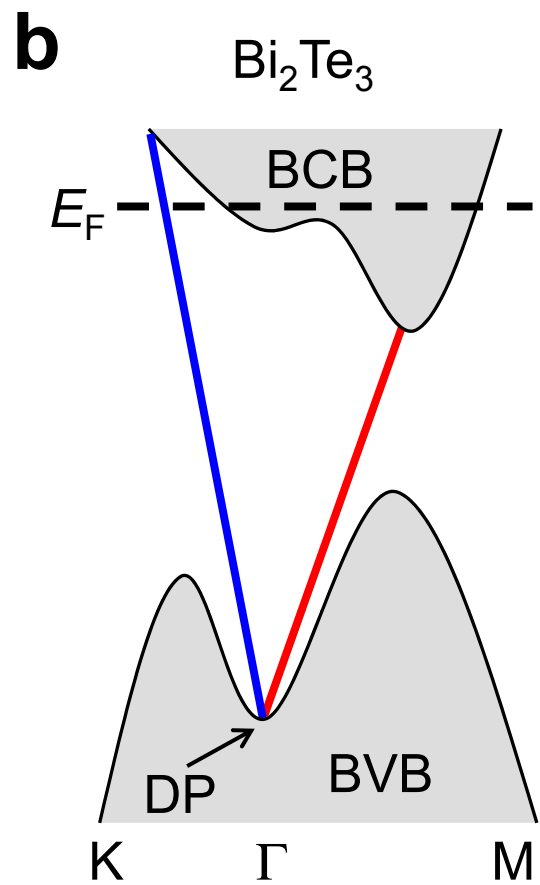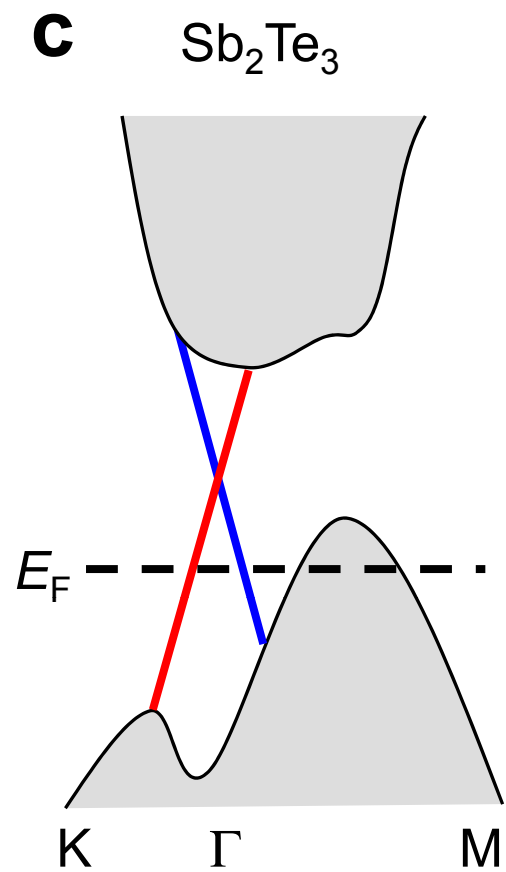

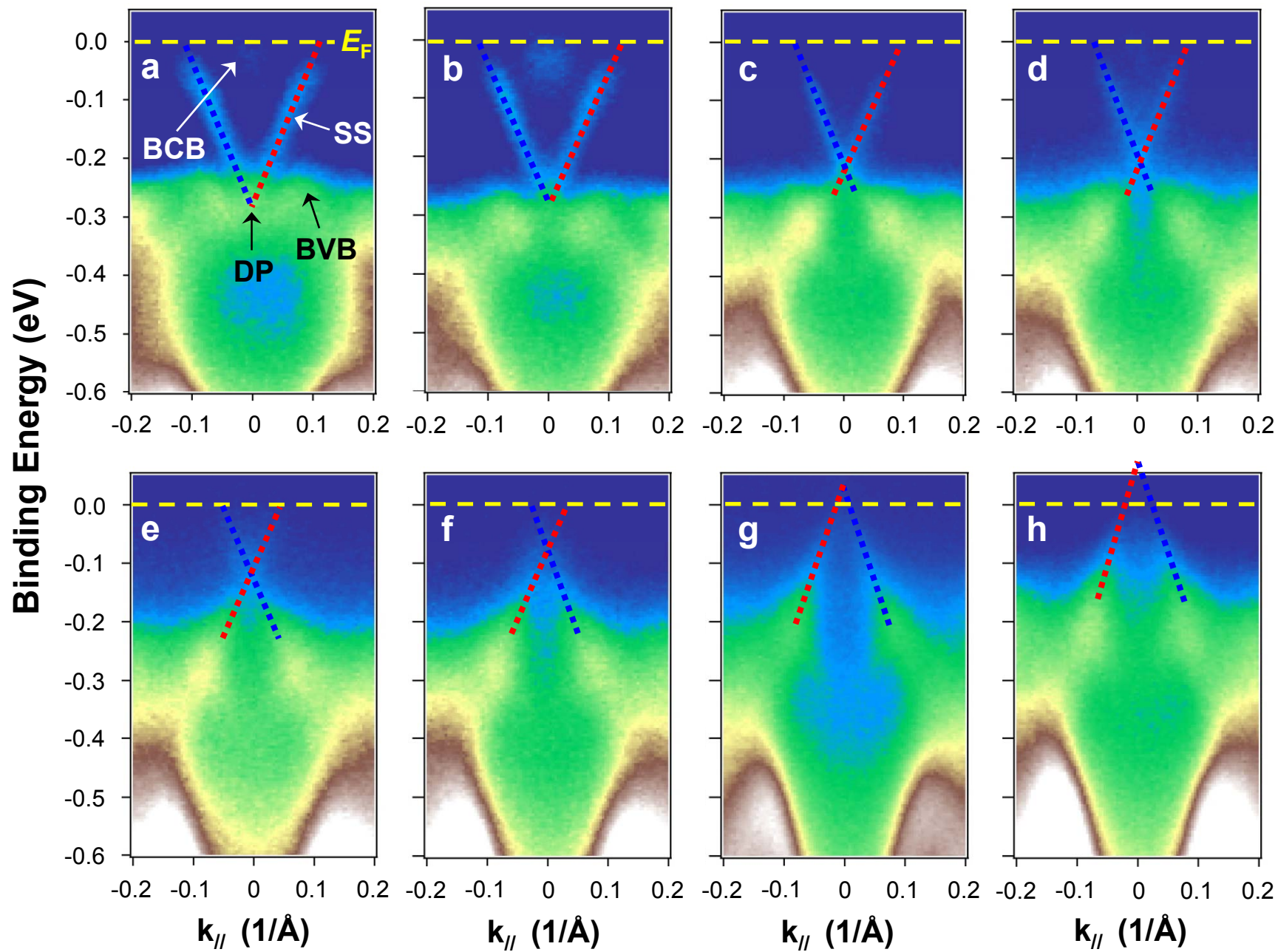

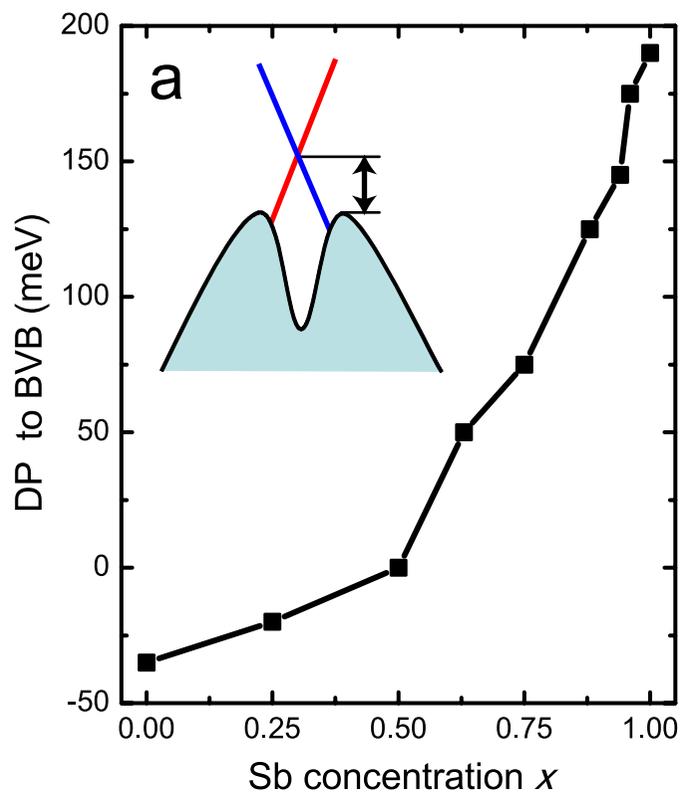 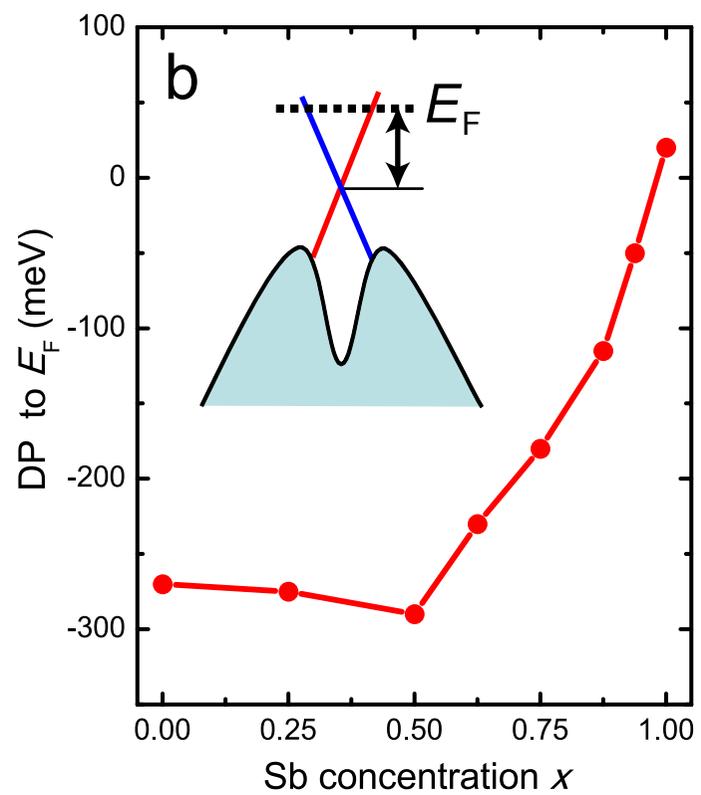 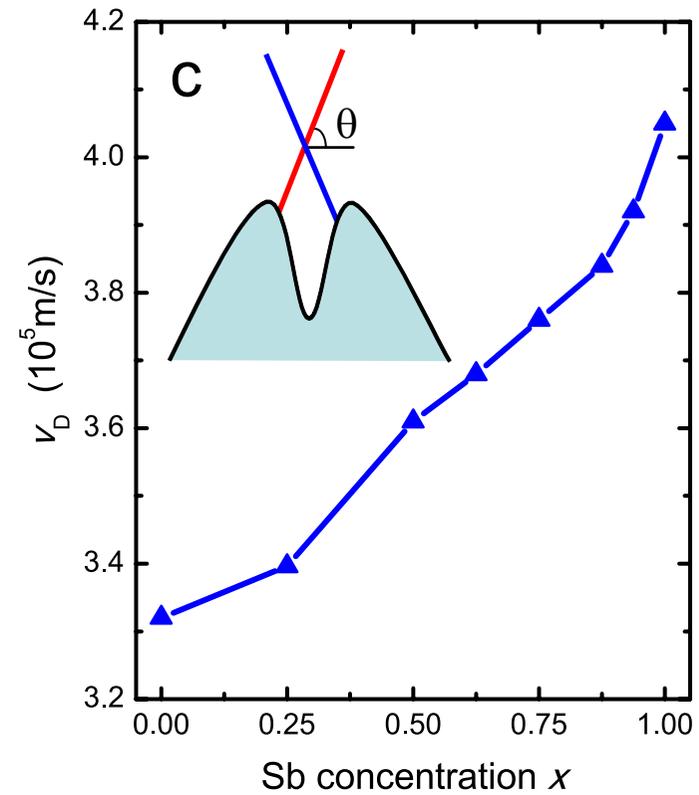

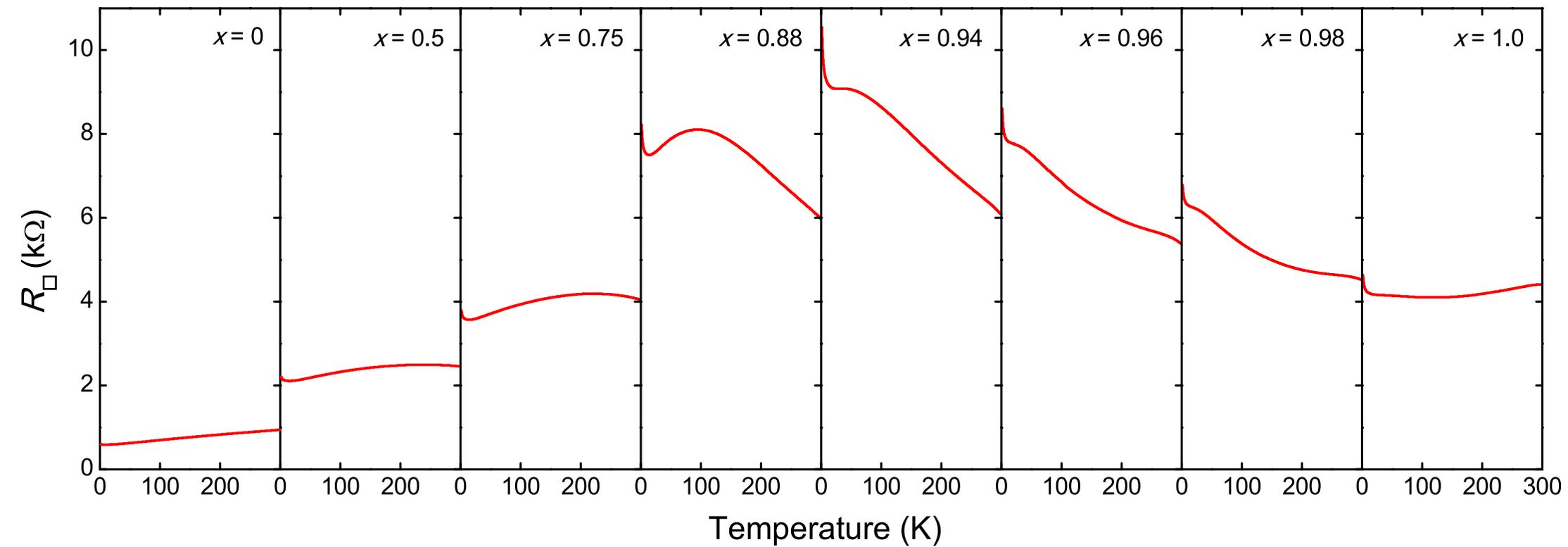

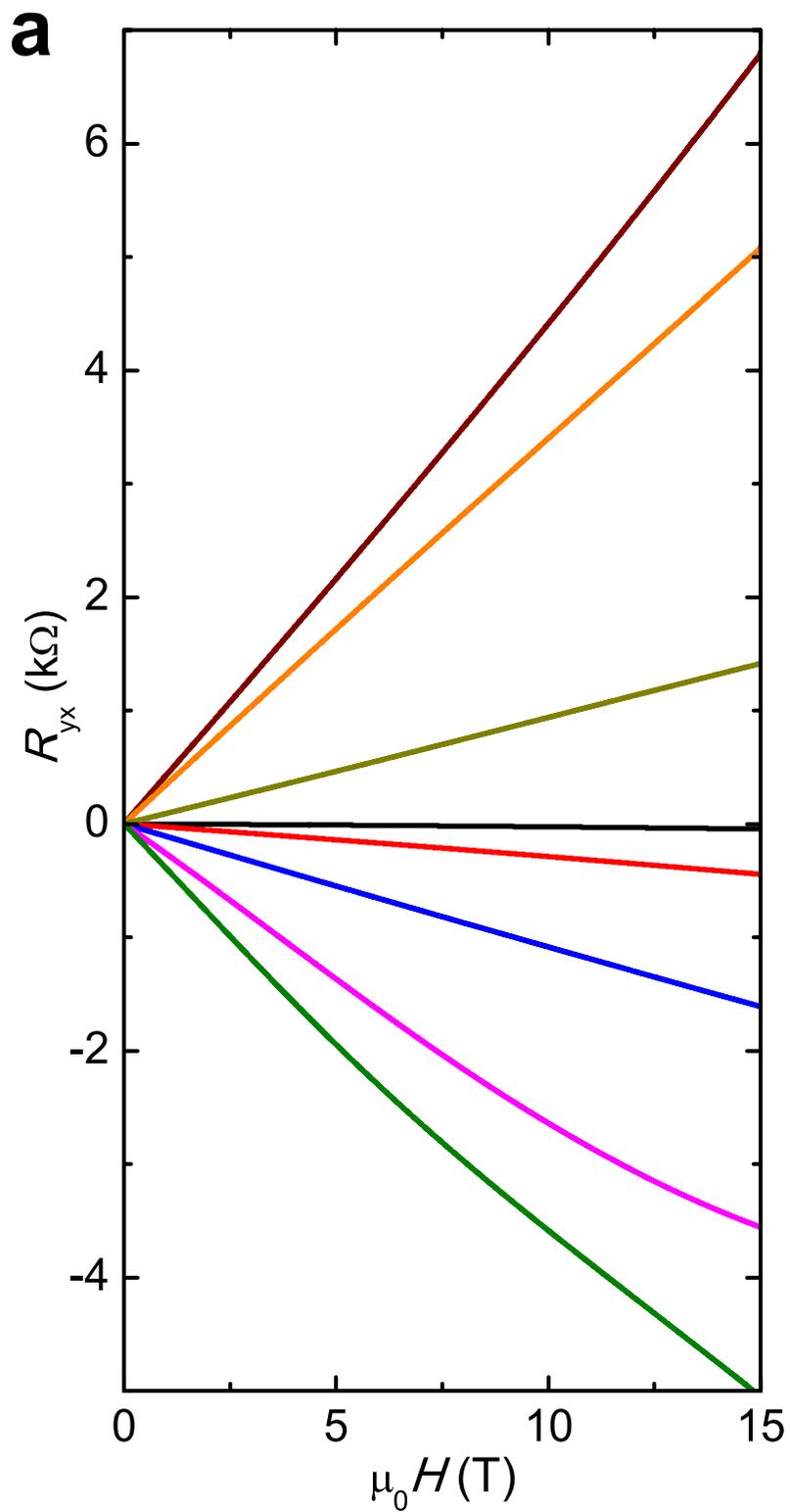
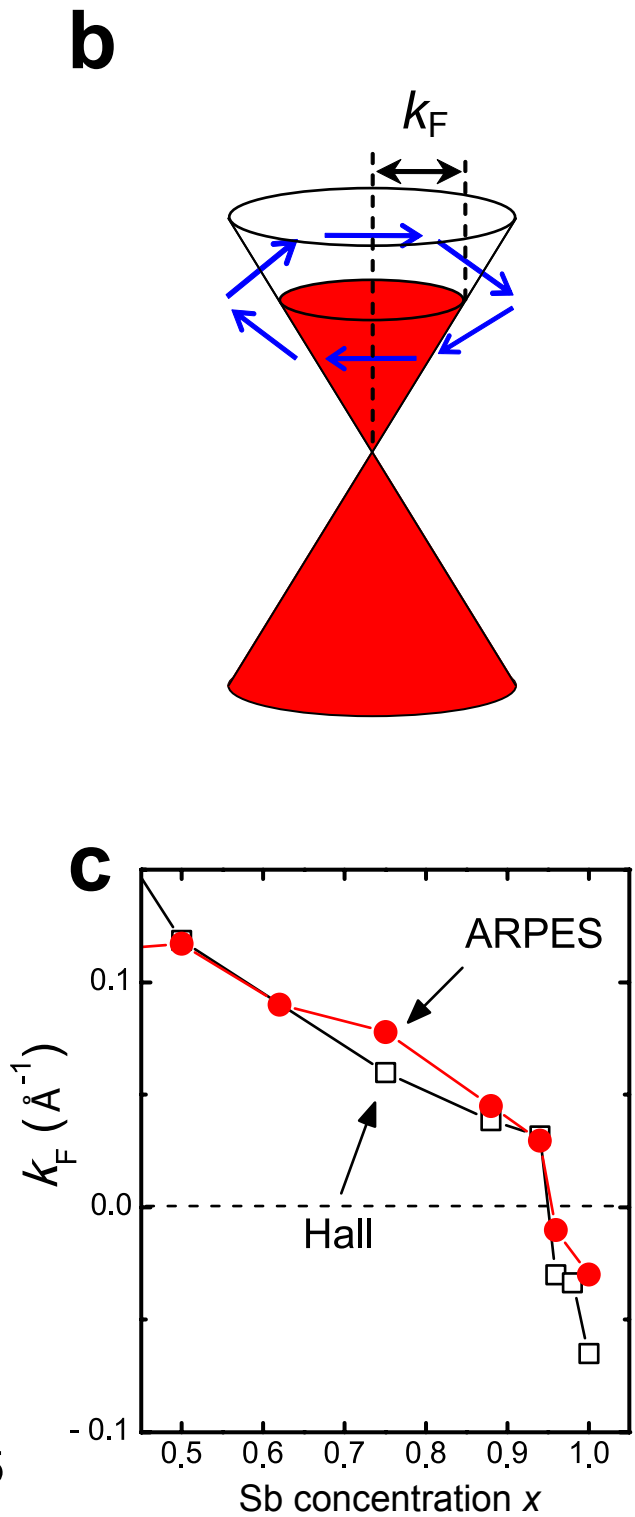
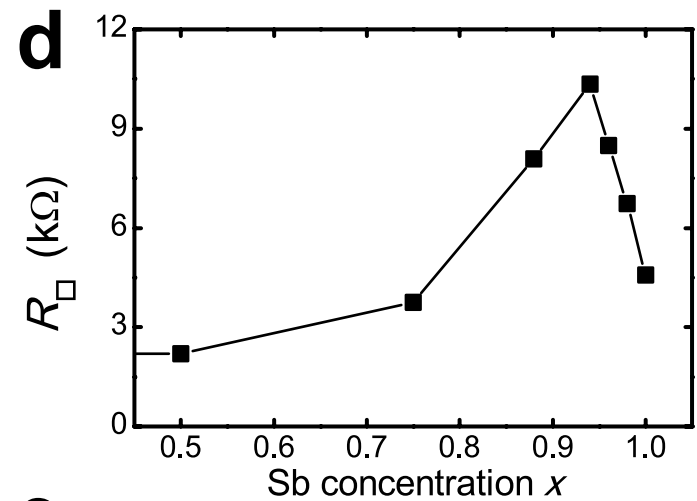
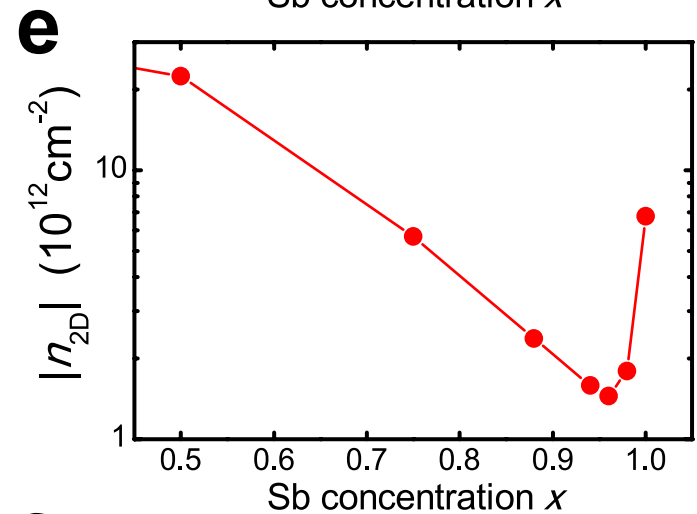
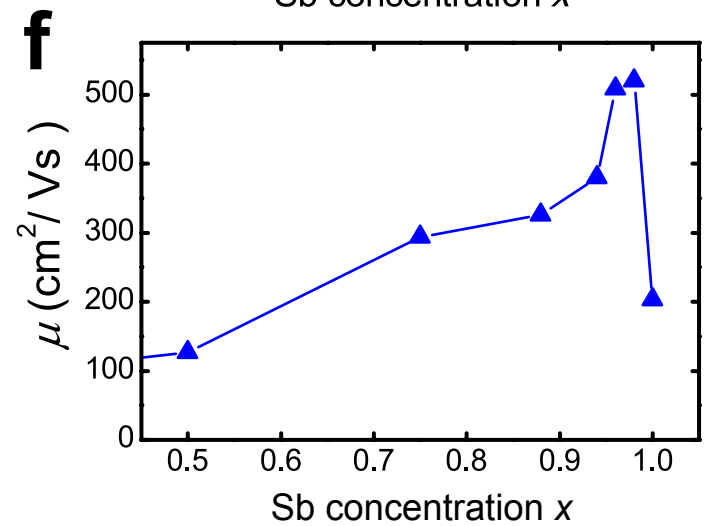

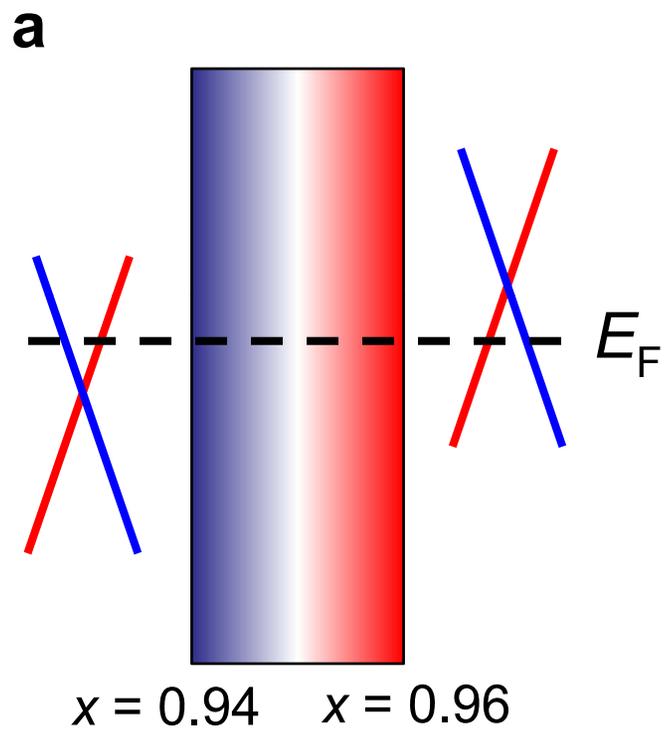 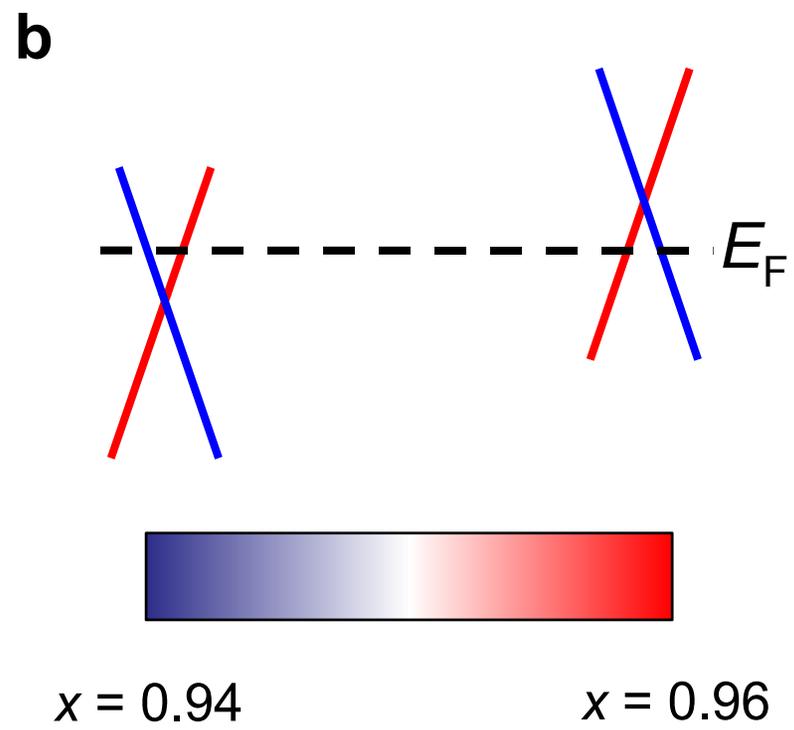

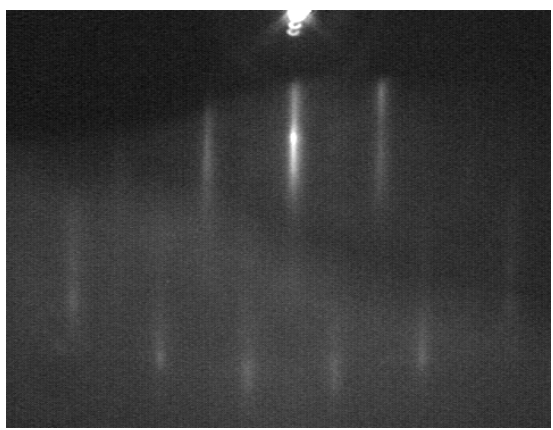

**Supplementary Figure S1 The *in situ* RHEED pattern of the MBE grown film.** The 1×1 RHEED (Reflection High Energy Electron Diffraction) pattern measured on the 5QL $(Bi_{0.12}Sb_{0.88})_2Te_3$ film. The sharpness of the pattern indicates the high quality of the sample.

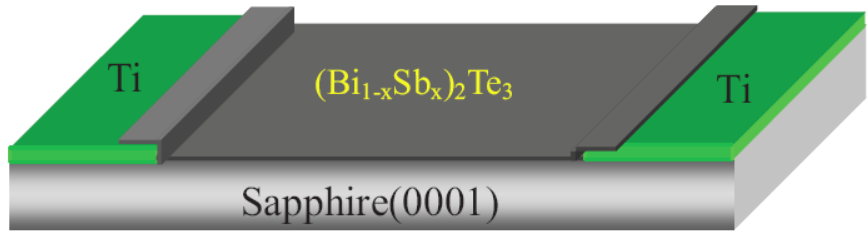

**Supplementary Figure S2 Sketch of the sample setup for *in situ* ARPES measurement.** To avoid sample charging during ARPES measurements due to the insulating sapphire substrate, a 300-nm-thick titanium (Ti) film is deposited at both ends of the substrate which is connected to the sample holder. The sample is grounded through these contacts once a continuous film is formed.

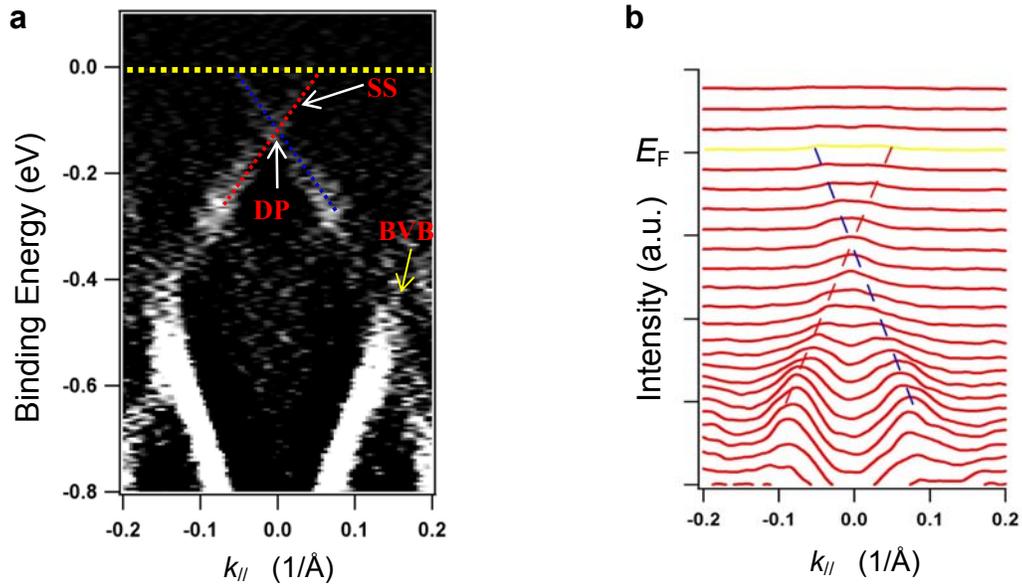

**Supplementary Figure S3 Differential ARPES spectrum and MDC of the *x* = 0.88 film.** **a**, Second-order differential ARPES spectrum of the 5QL $(Bi_{0.12}Sb_{0.88})_2Te_3$ film along the K-Γ-K direction. The surface states (SS), Dirac point (DP) and bulk valence band (BVB) are marked. **b**, The momentum distribution curve (MDC) measured on the same sample. The velocity of the Dirac fermions ($v_D$) and the Fermi wavevector ($k_F$) can be determined reliably when the peak positions of the surface states are obtained by fitting the MDCs with Lorentz functions.

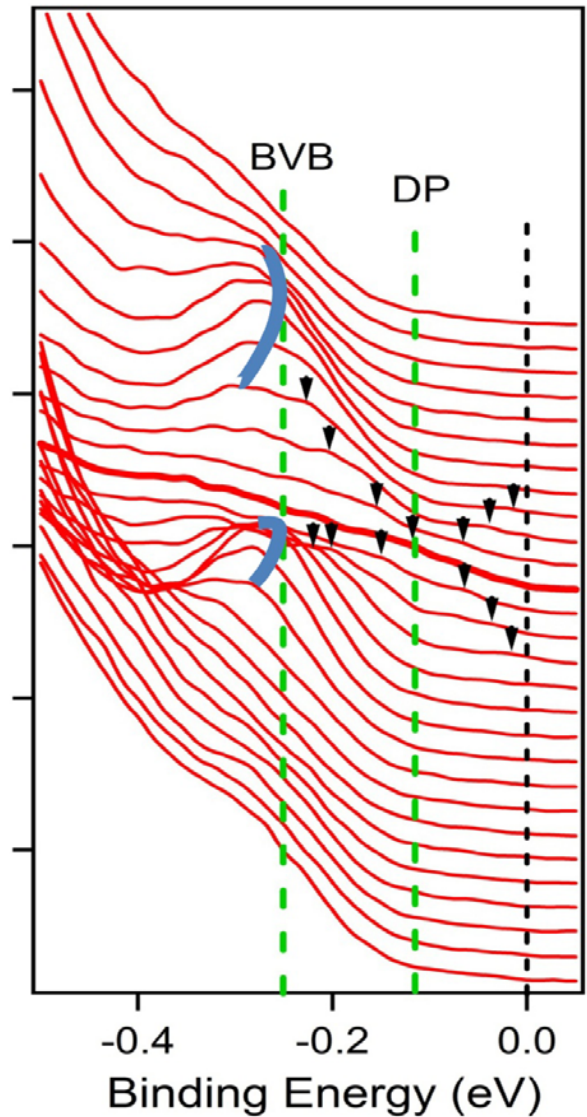

**Supplementary Figure S4 EDC of the *x* = 0.88 film measured by ARPES.** The energy distribution curve (EDC) measured on the 5QL $(Bi_{0.12}Sb_{0.88})_2Te_3$ film. The positions of DP and the top of BVB near Γ point are about -115 meV and -245 meV, respectively. So the distance from DP to the top of the BVB is around 130meV.

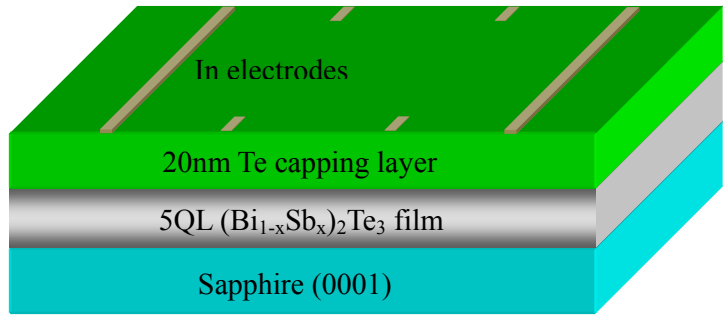

**Supplementary Figure S5 Schematic setup of the MBE grown film for transport measurements.** The thin film is grown on insulating sapphire (0001) substrate and then covered by 20nm of amorphous Te capping layer before it is taken out of the UHV chamber for *ex situ* transport measurements. The electrical contacts are made by mechanically pressed indium electrodes. The average size of the sample is about 2mm by 5mm. The thickness in the figure is not to scale.

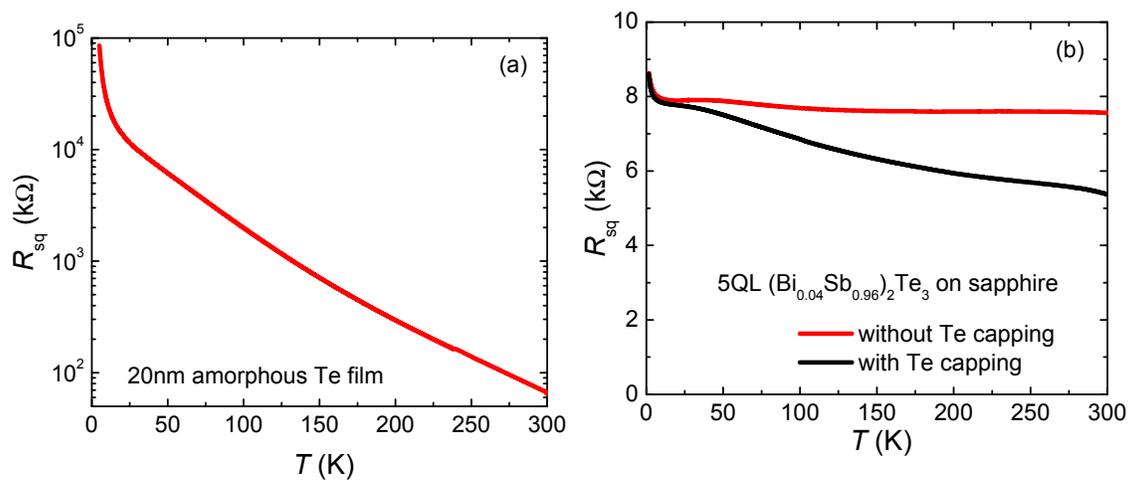

**Supplementary Figure S6 Transport studies on the effect of the amorphous Te capping layer. a**, The temperature evolution of the sheet resistance of the 20nm amorphous Te film grown on sapphire substrate showing a semiconductor behavior. The resistance values are much larger than that of the TI films. **b**, The temperature dependent resistance of the 5QL $(Bi_{0.04}Sb_{0.96})_2Te_3$ film covered with Te capping layer (black) compared with the one without Te capping (red). The transport properties of the films are qualitatively the same.

## Supplementary Methods

**The effect of film thickness on the transport properties**

All the data shown in this work are taken on $(Bi_{1-x}Sb_x)_2Te_3$ films with thickness $d = 5QL$. There are two main considerations for choosing this particular film thickness. The first requirement is that the TI film must be thick enough so that there is no coupling between the top and bottom surfaces. For the $(Bi_{1-x}Sb_x)_2Te_3$ compounds, the coupling between the two surfaces vanishes for $d \geq 4QL$. The 5QL films used here have completely decoupled surface states, as seen by the absence of energy gap at the Dirac point measured by ARPES. Our collaborators also performed scanning tunneling spectroscopy, which have much better energy resolution than ARPES, on similar films and confirmed this conclusion. The second requirement concerning the film thickness is that it should be thin enough so that the bulk conduction can be safely ignored. Although we can tune the Fermi level into the bulk energy gap by using the band engineering technique, there is always unavoidable bulk conduction via variable range hopping through the defects. With increasing film thickness this contribution becomes more pronounced. Take the $Bi_2Te_2Se$ TI single crystal as an example. Even though the Fermi level lies in the bulk energy gap and the bulk defect density is very low, the bulk states still account for ~94% of the total conduction[29]. A major advantage of the MBE technique is that we can grow ultrathin films of TIs with large surface to volume ratio. In the 5QL films the bulk conduction is negligible so that the transport is dominated by surface states. Therefore the 5QL thickness is chosen because it is the optimal thickness where the coupling of the surfaces vanishes and the bulk conduction can be safely ignored. The ability to reach the ultrathin regime of TIs in a controlled manner is actually an important advantage of the MBE growth technique.

**Analysis of the ARPES data**

In Figures 3a to 3c of the main text we show the $x$ dependence of the relevant parameters of the surface band structure measured by ARPES. Here we use the spectrum of the $x = 0.88$ film as an example to demonstrate the ARPES data analysis procedure.

In order to enhance the visibility of the band structure, the second-order differential

ARPES spectrum of the 5QL (Bi$_{0.12}$Sb$_{0.88}$)$_2$Te$_3$ film is shown in Supplementary Figure S3a. The surface states with Dirac-like dispersion can be seen clearly and are marked by the dashed lines (the red and blue color indicates the different spin polarization). The point where the two dashed lines meet is defined as the Dirac point (DP). The energy difference between the DP and the Fermi level $E_F$ (yellow dashed line at zero binding energy) is the quantity plotted in Figure 3b of the main text.

Supplementary Figure S3b shows the corresponding momentum distribution curves (MDC). The peak positions of the Dirac surface states are obtained by fitting the MDCs with Lorentz functions. The velocity of the Dirac fermions, $v_D = 3.8 \times 10^5$ m/s, is then derived from the linear $E$ vs $k$ dispersion near the DP. The Fermi wavevector $k_F$ of the surface states can be read from the $k$ value of the crossing point between the surface states and the $E_F$ and is found to be $k_F = 0.057$ Å$^{-1}$ for the $x = 0.88$ sample. The $v_D$ and $k_F$ values are plotted in Figure 3c and Figure 5c of the main text respectively.

Supplementary Figure S4 shows the energy distribution curves (EDC) of the 5QL (Bi$_{0.12}$Sb$_{0.88}$)$_2$Te$_3$ film. The peaks indicated with black arrows are the Dirac surface states. The prominent and broad features at around $E_B = -245$ meV (indicated with thick blue lines) result from the hybridization between surface states and bulk band[21,30]. So the features can be regarded as the top of bulk valence band (BVB) near Γ point. In this sample, the positions of DP and the top of BVB near Γ point are about -115 meV and -245 meV, respectively. So the distance from DP to BVB is 130meV, as plotted in Figure 3a of the main text.

**Determination of the $x$ value in (Bi$_{1-x}$Sb$_x$)$_2$Te$_3$**

The $x$ values in the (Bi$_{1-x}$Sb$_x$)$_2$Te$_3$ films are determined through two independent methods. The first method is to calculate it through the Bi/Sb flux ratio used in the growth of a (Bi$_{1-x}$Sb$_x$)$_2$Te$_3$ film. In the MBE growth of pure Bi$_2$Te$_3$ or Sb$_2$Te$_3$, the substrate is kept in a Te-rich environment and the growth rate is only determined by the flux of Bi or Sb[29]. The flux of Bi or Sb at a given source temperature can be accurately calibrated by the growth rate of pure Bi$_2$Te$_3$ or Sb$_2$Te$_3$ films measured by RHEED oscillation and quantum well states revealed by ARPES[30,31]. Therefore in the MBE-grown (Bi$_{1-x}$Sb$_x$)$_2$Te$_3$ films $x$ can be determined by the calibrated flux ratio between Bi and Sb. To vary the $x$ value in the $(Bi_{1-x}Sb_x)_2Te_3$ films, we control the source temperatures of Bi and Sb. Recall that in MBE growth the flux ($J$) and the source temperature ($T$) is related by $J \propto P/\sqrt{T}$, where $P$ is the vapor pressure at the source temperature $T$[32]. By choosing different source temperatures of Bi and Sb, hence the flux ratio between them, we can systematically tune the $x$ value of the film. For the $x = 0.88$ film, for example, the temperatures of the sources are $T_{Bi} = 505°C$, $T_{Sb} = 365°C$, $T_{Te} = 325°C$, and the temperature of the substrate is $T_{substrate} = 180°C$. The growth rate under these conditions is about 0.15 QL/min.

In the second method we directly measure the composition of the $(Bi_{1-x}Sb_x)_2Te_3$ film by using inductively coupled plasma atomic emission spectroscopy (ICP-AES). For ICP-AES measurements, thicker films (above 50nm) are needed to give sufficient accuracy. Growth of such thick films is rather time-consuming with the current growth rate of our MBE system. Therefore we did not perform the ICP-AES measurement on every sample. In the two $(Bi_{1-x}Sb_x)_2Te_3$ samples that we have measured, the difference between the $x$ value determined by ICP-AES and that estimated from the flux ratio is around 5%. The reasonably good agreement indicates that the $x$ value estimated from the Bi/Sb flux ratio, which is the number used in the manuscript, is quite reliable.

**Effect of the Te capping layer on the transport properties of the $(Bi_{1-x}Sb_x)_2Te_3$ films**

For the device structure that we are using, the total measured conductance consists of three terms. In addition to that from the TI film, the conductance of the substrate and the capping layer is also picked up in the measurements. The sapphire substrate is an excellent electrical insulator and its contribution to electrical conduction is negligible. The amorphous Te capping material is a semiconductor with energy gap larger than 300meV. With a thickness of 20nm, the Te capping layer should have very small contribution to the transport results. To justify this point, we have measured the resistance of a 20nm amorphous Te film grown on sapphire and contacted by In electrodes. Supplementary Figure S6a displays the temperature evolution of the sheet resistance of the Te film ($R_{Te}$), which shows an insulating behavior over the whole temperature range. The $R_{Te}$ value at room temperature is around 70kΩ, more than 10 times larger than the total resistance of the most resistive $(Bi_{1-x}Sb_x)_2Te_3$ film. Therefore at room temperature the error in the sample resistance value induced by the parallel conduction of the Te capping layer is less than 10%. The $R_{Te}$ value increases rapidly with lowering $T$. It reaches 1MΩ at $T = 120$K and at the base temperature $T = 1.5$K it is close to 100MΩ, which is about four orders of magnitude larger than that in the most resistive $x = 0.94$ film. Therefore the contribution of the Te film to the measured transport properties can be safely ignored, especially at the most interesting low $T$ regime where the Hall effect is taken.

Another concern about the Te capping layer is whether it will cause any significant change of the surface electronic structure of the TI films. To clarify this issue, we have measured the transport properties of a 5QL $(Bi_{0.04}Sb_{0.96})_2Te_3$ film without Te capping. During the sample mounting process particular care was taken to minimize the exposure of the film to ambient air. As shown in Supplementary Figure S6b, at low $T$ the resistance value and the diverging insulating behavior of the uncapped film is highly consistent with that of the Te-capped $x = 0.96$ film shown in the main text. At high $T$ there are substantial deviations between the two curves, where the uncapped film shows a larger resistance and weaker $T$ dependence than the capped film. The difference is too large to be explained merely by the parallel conduction from the 20nm amorphous Te layer in the capped film. We believe that the larger resistance and weaker $T$ dependence in the uncapped film is mainly due to the surface adsorbates that tends to degrade the quality of the surface and lower the mobility of the surface state Dirac fermions. Nevertheless, the qualitative behavior of the two films is very similar, suggesting that the Te capping layer does not significantly alter the electronic structure of the $(Bi_{1-x}Sb_x)_2Te_3$ films.

## Supplementary References